\documentclass[twoside,leqno,twocolumn]{article}

\usepackage[letterpaper]{geometry}

\usepackage{ltexpprt}

\usepackage{times}
\usepackage{helvet}
\usepackage{courier}
\usepackage{amsmath}
\usepackage{graphics}
\usepackage{graphicx}
\usepackage{xcolor}
\usepackage{authblk}

\graphicspath{ {Figures/} }

\begin{document} \fontsize{9.7}{11}\rm

\title{\Large Multi-Channel Sequential Behavior Networks for User Modeling in Online Advertising}
\author{Iyad Batal}
\author{Akshay Soni}
\affil{Microsoft Bing Ads}
\date{}

\maketitle



\begin{abstract}

Multiple content providers rely on native advertisement for revenue by placing ads within the organic content of their pages. We refer to this setting as ``queryless'' to differentiate from search advertisement where a user submits a search query and gets back related ads. Understanding user intent is critical because relevant ads improve user experience and increase the likelihood of delivering clicks that have value to our advertisers. 

This paper presents Multi-Channel Sequential Behavior Network (MC-SBN), a deep learning approach for embedding users and ads in a semantic space in which relevance can be evaluated. Our proposed user encoder architecture summarizes user activities from multiple input channels--such as previous search queries, visited pages, or clicked ads--into a user vector. It uses multiple RNNs to encode sequences of event sessions from the different channels and then applies an attention mechanism to create the user representation. A key property of our approach is that user vectors can be maintained and updated incrementally, which makes it feasible to be deployed for large-scale serving. We conduct extensive experiments on real-world datasets. The results demonstrate that MC-SBN can improve the ranking of relevant ads and boost the performance of both click prediction and conversion prediction in the queryless native advertising setting.

\end{abstract}

\section{Introduction}

Online advertising is a multibillion-dollar market that provides substantial revenue for search engines and content providers. Search advertising (SA) places ads on the result page of a search engine and is arguably the most profitable business model on the web today. In this setting, advertisers bid on keywords for their ads. When the user enters a search query, it is matched against the bid keywords and relevant ads are allocated to available ad slots.

Another form of online advertising that is gaining popularity is \emph{native} advertising--also called contextual advertising--that directly places ads on third-party web pages, matching in form the organic content of the page. In this setting, the ad platform acts as a mediator between the advertiser, the publisher, and the user with the goal of increasing revenue (shared between the platform and the publisher) and improving user experience. We refer to this as ``queryless advertising" since users do not explicitly provide search queries as in SA.

Understanding user intent is critical because increasing ads relevance improves user experience and also increases the probability of high-quality clicks\footnote{Clicks that have business value to advertisers and can potentially lead to conversions.}\cite{Wang:2002}. This is usually easier in SA because the user explicitly and concisely expresses her current intent in form of a search query. On the other hand, delivering relevant and personalized ads is more challenging in queryless advertising because inferring the exact user intent is a complex task, given the heterogeneity of user activities and difficulty of estimating motives from indirect sources.

Deep learning (DL) models have been successfully applied for learning useful semantic embeddings for queries and documents \cite{Huang:2013,Shen:2014,Zhai:2016}. Such models can be used for ranking ads (or documents) with respect to a specific \emph{search query}, as in the SA setting. We extend such techniques to the queryless setting by learning embeddings that are useful for ranking ads with respect to a specific \emph{user state} at a given time, which is more challenging to represent and model compared to a search query.

This paper presents the Multi-Channel Sequential Behavior Network (MC-SBN) model. The model consists of a user encoder and an ad encoder, which are jointly trained from logs to learn user preferences. MC-SBN is designed to be practical for large-scale and latency-sensitive serving systems. It is deployed in our production system which serves tens of millions of users, processes hundreds of thousands of user events per seconds, while ensures that serving latency is within tens of milliseconds. A key property is that our user encoder network allows incremental updates of user representations, and hence these representations can be updated offline as events stream through the system. 

Our contributions are summarized as follows:
\begin{itemize}
\item We propose a flexible user encoder architecture that consumes user activities from multiple heterogeneous sources of data (e.g., user's previously submitted search queries, visited pages, or clicked ads) and summarizes them as a user intent vector. This network processes time-segmented sessions of events from each source (channel) separately via a GRU-based RNN. The channel-specific representations are then combined using an attention mechanism to perform credit assignment among channels based on intent importance.
\vspace{-0.03in}
\item We present a learning formulation to train the model by utilizing search advertising logs for alleviating data selection bias \cite{schnabel2012}. 
\vspace{-0.03in}
\item The MC-SBN model—trained end-to-end from logs—is designed to decompose into a user encoder and a separate ad encoder that can run independently of each other in production to minimize serving latency.
\vspace{-0.03in}
\item We present large-scale experiments on real-world datasets obtained from the Bing advertising system. We compare MC-SBN with several existing state-of-the-art models to demonstrate the benefits of the proposed model, as well as to highlight the key differences in the problem setup assumed by us and the other approaches. We also present an ablation study to understand the benefits provided by the different components of our model. Lastly, we present experiments for click and conversion predictions in the queryless native advertising setting.

\end{itemize}

The rest of the paper is organized as follows. Section \ref{sec:related} outlines the related research. Section \ref{sec:proposed} describes the MC-SBN model. We describe the learning formulation (Section \ref{sec:learning-formulation}), present details of the user encoder network (Section \ref{sec:user-network}), and illustrate how the model is deployed in production (Section \ref{sec:system}). Section \ref{sec:experiments} presents the experimental evaluation. Lastly, Section \ref{sec:conclusion} concludes the paper.

\section{Related Work}
\label{sec:related}

\subsection{Semantic Matching in IR}
There has been a lot of research work in the information retrieval (IR) field concerning the problem of matching search queries to relevant documents (or ads) on a semantic level. These models aim to address the limitation of lexical matching by first representing a query and a document as two vectors in a lower-dimensional semantic space and then computing similarity in that space. 

Earlier approaches on semantic matching, such as \cite{Hofmann:1999}, \cite{Blei:2003}, and \cite{Le:2014}, were mostly trained in an unsupervised manner using objective functions that are only loosely coupled with the actual evaluation metric of the retrieval task. Later on, a series of DL-based models \cite{Huang:2013,Shen:2014,Zhai:2016} have been introduced to discriminatively train the embeddings using click-through logs, which are abundant in search engines. These models are trained to maximize the conditional likelihood of clicked documents given a query, which is often approximated using negative sampling \cite{Mikolov:2013}. 

The above-mentioned models are useful for ranking ads with respect to a specific search query (few keywords the user provides to specify her current intent), as in the search advertising scenario. Our work extends these models by learning to rank ads with respect to a user state, which is a more difficult entity to represent and model as compared to a search query.

\vspace{-0.02in}
\subsection{User Modeling}

There has been a growing interest in applying DL techniques to model user activities in recommender systems \cite{Zhang:2019}. RNNs were found particularly suitable for session-based recommendation, in which the task is to predict the next item a user will interact with based on his/her previous items. GRU2Rec \cite{Hidasi:2015} represents item sequences using a GRU-based RNN. Loyola et al. \cite{Loyola:2017} describe an encoder-decoder architecture for learning source-to-target transitions between pairs of items. Smirnova et al. \cite{Smirnova:2017} design a context-aware model using conditional RNN, which allows injecting context information (e.g., time of day) into the input and output layers of the RNN. Donkers et al. \cite{Donkers:2017} present a novel type of GRU that explicitly models individual users--using one-hot encoding of user ID--for improving next item recommendation. These approaches represent the input of the RNN as a sequence of item IDs (i.e., by learning ID-based embedding of items). As we will show later in our experiments, this is not ideal in our domain because new ads or pages are constantly created, and such representation cannot properly generalize to new or tail events. 

DL-based user models have also been explored in other recommendation domains. Covington et al. \cite{Covington:2016} describe the movie recommendation system of YouTube. Their model represents a user by the average ID-based embedding of previous events (e.g., previously watched videos). This work was extended in \cite{Beutel:2018} by using an RNN to model user watches and also by incorporating context information such as location or device type. Okura et al. \cite{Okura:2017} presents a news recommendation system using denoising autoencoders with weak supervision. Tang et. al. \cite{Tang:2017} describe an event recommendation model for social networks, which represents a user as a text document and an unordered list of categorical feature-value pairs.

Other recent papers propose applying attention to adaptively create user representations for e-commerce applications \cite{Zhou:2018, Ni:2018, Zhou:2019:DIEN, Yu:2019:long-short}. The idea is to maintain, for each user, the embeddings of the most recent $K$ activities. When scoring a specific candidate \emph{at runtime}, the model uses attention to determine which activities should be emphasized (or neglected) according to the candidate. In contrast to e-commerce applications, many of our user events might have little value for commercial targeting (such as informational queries or uncommercial page visits). Therefore, applying this approach effectively at scale poses challenges because it would require storing and processing a large number of user events for online serving. On the other hand, MC-SBN maintains a single vector per user, which is incrementally updated--by an independent offline process--based on the activities of the user.

\section{MC-SBN}
\label{sec:proposed}

The main objective of this work is to evaluate the relevance of a specific ad $a \in A$ to a specific user $u \in U$ at a given time $t$, and use that for ranking native ads. To achieve this, the model should be able to understand user intent from the observed history of activities and the content of the ad from its description.

User activities form a time-sorted sequence of events $\{e_1, e_2, ..., e_N\}$, where $e_i$ denotes the $i^{th}$ user event. Each event $e_i$ is described as a triplet $(t_i, c_i, v_i)$, where $t_i$ is the timestamp, $c_i$ is the channel/source (e.g., whether the event is a browsed page or a search query), and $v_i$ is the text associated with the event (e.g., the title of the page or keywords of the query).

\begin{figure}
\includegraphics[width=3.3in, height=1.9in]{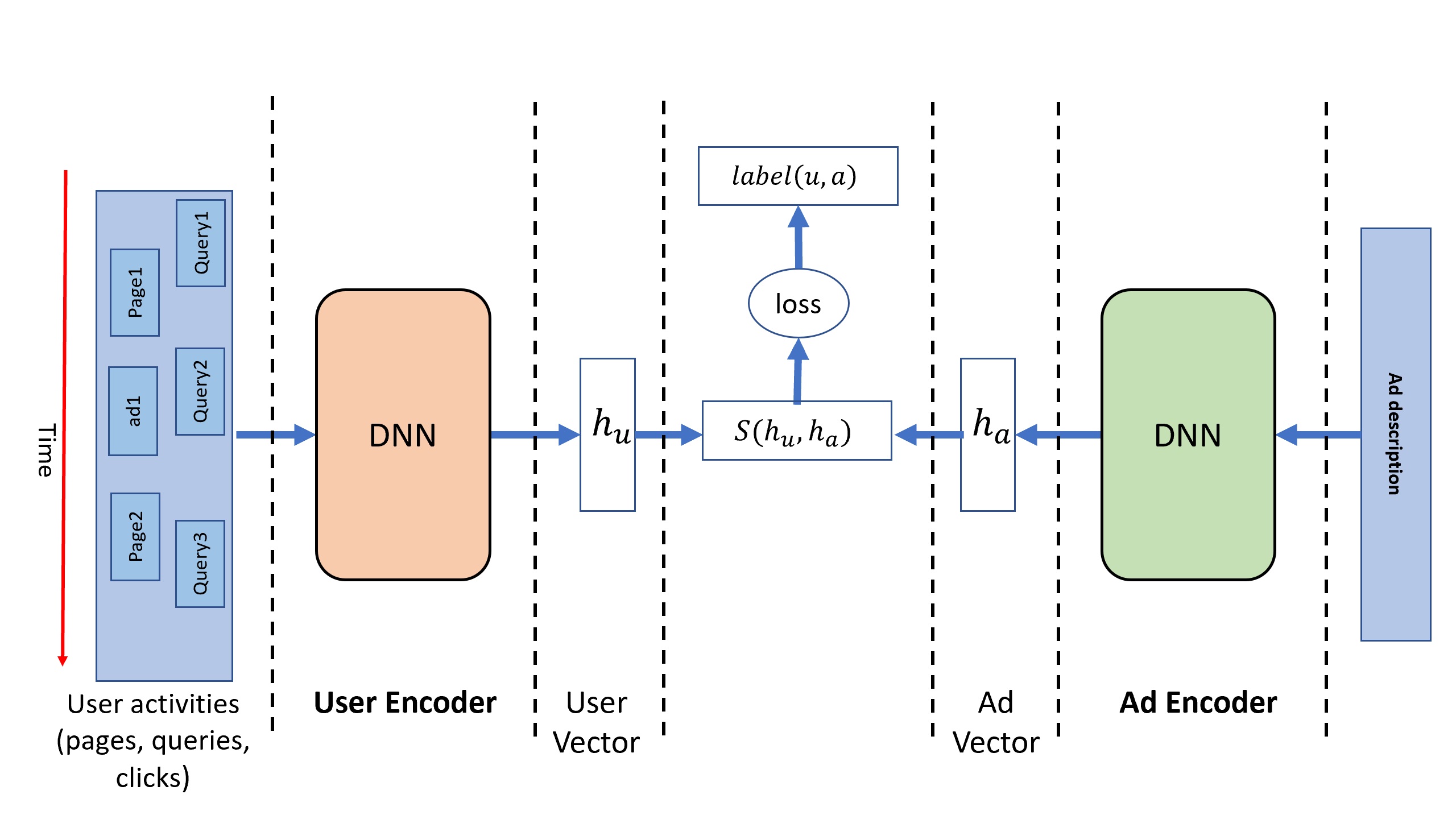}
\vspace{-0.1in}
\caption{\small{The architecture of MC-SBN that consists of a user encoder (left) and an ad encoder (right) that are jointly trained.}}
\label{fig:model_architecture}
\end{figure}

This paper presents, MC-SBN, an embedding-based approach for learning how to encode a user and an ad as two vectors in a lower-dimensional semantic space where relevance can be measured (Figure \ref{fig:model_architecture}). More specifically, we use neural networks to learn two functions: 1) the \emph{user encoder} $F_{\text{\emph{user}}}\left( u(t) \right)$ that takes activities of user $u$ up to time $t$ and represents it as a user vector $h_{u(t)}$, and 2) the \emph{ad encoder} $F_{\text{\emph{ad}}} \left( a \right)$ that takes the ad text and represents it as an ad vector $h_a$.

A similarity score $S(h_{u(t)}, h_a)$ is then used to assess $u$'s interest in $a$. Consequently, the learning formulation optimizes $F_{\text{\emph{user}}}$ and $F_{\text{\emph{ad}}}$ in a way that makes ``relevant'' user-ad pairs (defined more rigorously in Section \ref{sec:learning-formulation}) more similar than irrelevant pairs. Here we simply use the dot product\footnote{It is possible to use a multilayer network to represent $S$, but that did not show gains in our experimental evaluation over the dot product.} for computing $S$ since it is fast to compute and also allows for efficient retrieval of top $k$ ads using general-purpose nearest neighbor search libraries \cite{Liu:2005}. 

There has been a lot of work on designing networks for learning text embeddings \cite{Huang:2013, Shen:2014, Zhai:2016, Bansal:2016}, which can be utilized to define the ad encoder network. We evaluated several approaches and converged on using a bi-directional GRU to learn the ad embedding from the sequence of tokens in the text \cite{Peters:2018}. On the other hand, learning the user encoder network is the more challenging task and the main focus of this work.

\vspace{-0.07in}

\vspace{-0.02in}
\subsection{The User Encoder Network}
\label{sec:user-network}

The desired user encoder should be able to consume events from all input channels and summarize them in a vector that represents the user's current interests. We define the following properties for the model to satisfy:

\begin{enumerate}
\item Represent user events from associated text content. An ID-based representation of events, as in \cite{ Zhou:2019:DIEN, Zhou:2018, Qu:2018:PNN, Beutel:2018, Ni:2018, Smirnova:2017, Donkers:2017, Loyola:2017, Hidasi:2015}, is impractical in our domain because the space of possible events is very large and constantly changing.
\vspace{-0.03in}
\item Understand the temporal order of events and learn how to incorporate the recency effect of the different semantic concepts. In sequential recommendation problems--like ours--the most recent events usually play more important role, but different concepts (interests) can have different underlying timespans. 
\vspace{-0.03in}
\item Allow the model to learn different representations for the different input channels. The reason is because the semantics of previous searches can be different from page visits as these may explain different aspects of the user behavior. Thus, combining all events and treating them as a single channel can result in loss of information. 
\vspace{-0.03in}
\item Maintain a concise representation of users that can be \emph{incrementally} updated to facilitate integration in large-scale serving systems. We avoided techniques that require storing and processing the entire user behavior at runtime scoring \cite{Zhou:2018, Ni:2018, Zhou:2019:DIEN, Yu:2019:long-short}.
\end{enumerate}

\begin{figure}
\includegraphics[width=3.2in, height=1.9in]{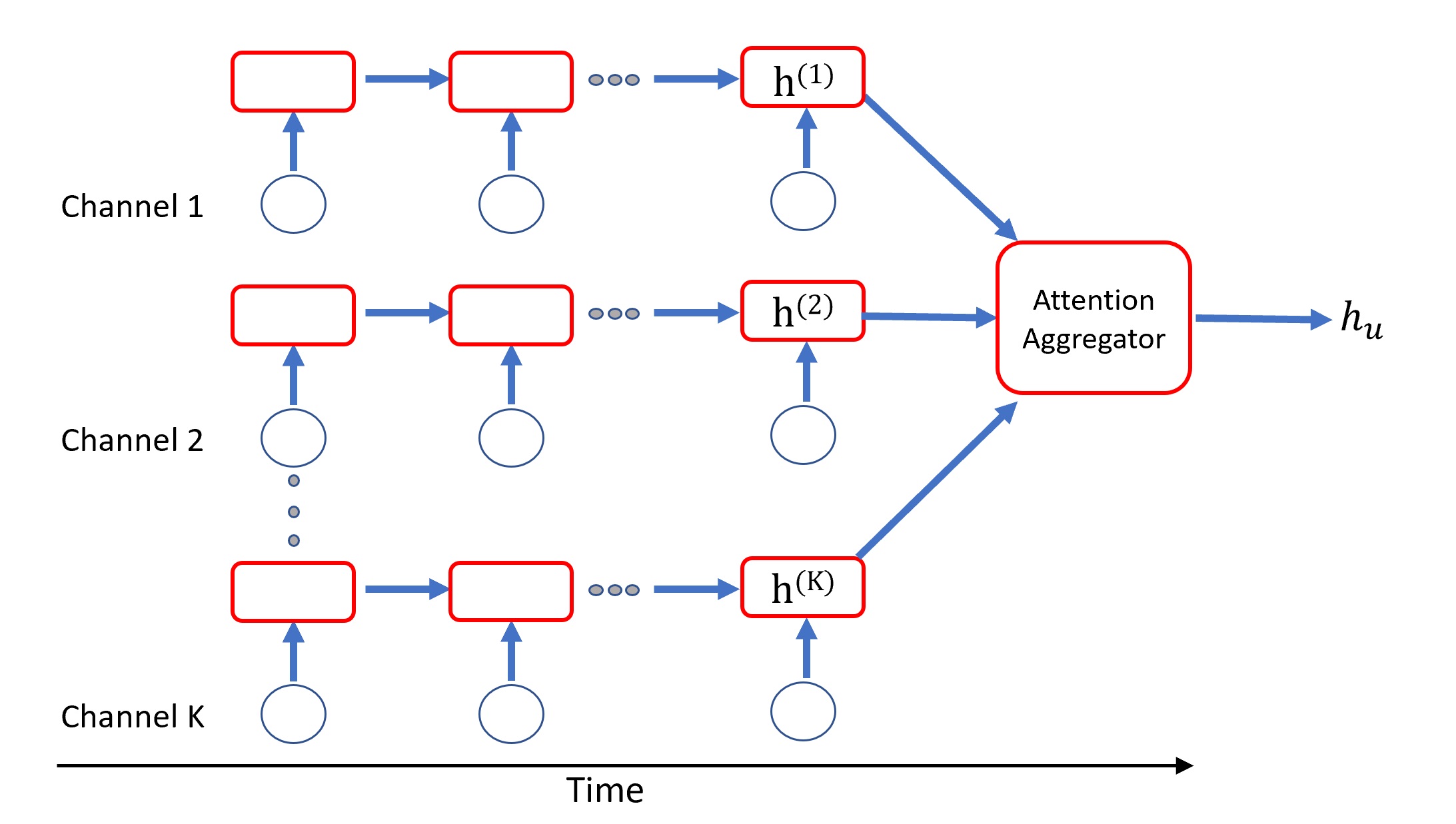}
\vspace{-0.1in}
\caption{The user encoder network.}
\label{fig:user-encoder}
\end{figure}

To achieve these properties, the proposed user encoder network uses multiple RNNs to encode the text content of events from each channel, and then applies an attention mechanism to create the final user representation (see Figure \ref{fig:user-encoder}). Once the entire MC-SBN model is trained end-to-end from logs, its user encoder network gets applied in an offline batch mode to keep updating user vectors as events are streamed through the system. Those user vectors are then used at serving time to efficiently score candidate ads.

\vspace{-0.07in}

\subsubsection{Input Representation}
\label{sec:channel-representation}

One approach to model input of a specific user channel is to feed the RNN a sequence of events, one at a time. For example, the ``page visits'' channel would consist of a sequence of most recent $T$ pages the user has browsed. The issue with this approach is that user data can be very skewed, where some users have a large number of events while others have much fewer. If we treat each event as a separate input of the RNN, the most recent $T$ events could span last one hour of history for an active user or could span several days for a less active user. Moreover, the sampling frequency of events from the different channels can be very different, even for the same user. For example, page visits generally have a higher observation rate compared to clicked ads. 

To create a more consistent representation of users, we propose segmenting events from each channel using fixed size intervals and providing to the RNN a \emph{sequence of sessions}, instead of a sequence of single events. In this way, each session is represented using the BoW (a multi-hot vector in the vocabulary space) of all events that co-occur in that same session.

\subsubsection{Sequential Modeling}
\label{sec:channel-modeling}

Let $\{ x^{(k)}_1, x^{(k)}_2, ..., x^{(k)}_T: x^{(k)}_t \in \{0,1\}^{|V|} \}$ denote the most recent $T$ sessions from the $k^{th}$ channel of a specific user; where $|V|$ is the size of the vocabulary and $x^{(k)}_i$ is the sparse multi-hot vector representation of the $i^{th}$ session. The hidden state at step $t$ gets updated according to the following recurrent formula:

\vspace{-0.06in}
$$h^{(k)}_t = f(x^{(k)}_t, h^{(k)}_{t-1})$$
\vspace{-0.04in}

We use a GRU-based RNN \cite{Cho:2014} to model the sequence of event sessions. A GRU network uses two special parameterized gates, the \emph{update gate} ($z_t$) and the \emph{reset gate} ($r_t$), which help the network to learn when and how to update the hidden state (i.e., the channel representation) when a new session of activities is observed. The new updated state is computed as follows: 

\vspace{-0.06in}
\begin{align}
\begin{split}
& c^{(k)}_t = \text{tanh} (W_h x^{(k)}_t + U_h (r_t \circ h^{(k)}_{t-1}) + b_h) \\
& h^{(k)}_t = (1-z_t) \circ h^{(k)}_{t-1} + z_t \circ c^{(k)}_t
\end{split}
\label{eq:gru-update}
\end{align}
\vspace{-0.04in}

The most recent state $h^{(k)}_T$ represents the \emph{current} content of the $k^{th}$ channel of a specific user. We abbreviate it as $h^{(k)}$ from hereafter to simplify the notation.

\subsubsection{Attention Aggregation}
\label{sec:attention}

The representations of individual channels $h^{(1)}, ..., ~h^{(K)}$ need to be combined to obtain the final user vector $h_u$ (see Figure \ref{fig:user-encoder}).

Average pooling and max pooling are commonly used aggregation methods. However, both cannot perform proper credit assignment among the individual channels. For example, consider a user whose page visits clearly indicate her commercial intent (e.g., browsing a specific product on an e-commerce site), but did not submit relevant queries through our search engine. If we use a fixed pooling technique, like average or max pooling, the important signal from the pages channel would be diluted with the less useful signal from the queries channel.

\begin{figure}[h]
\includegraphics[width=0.8\columnwidth]{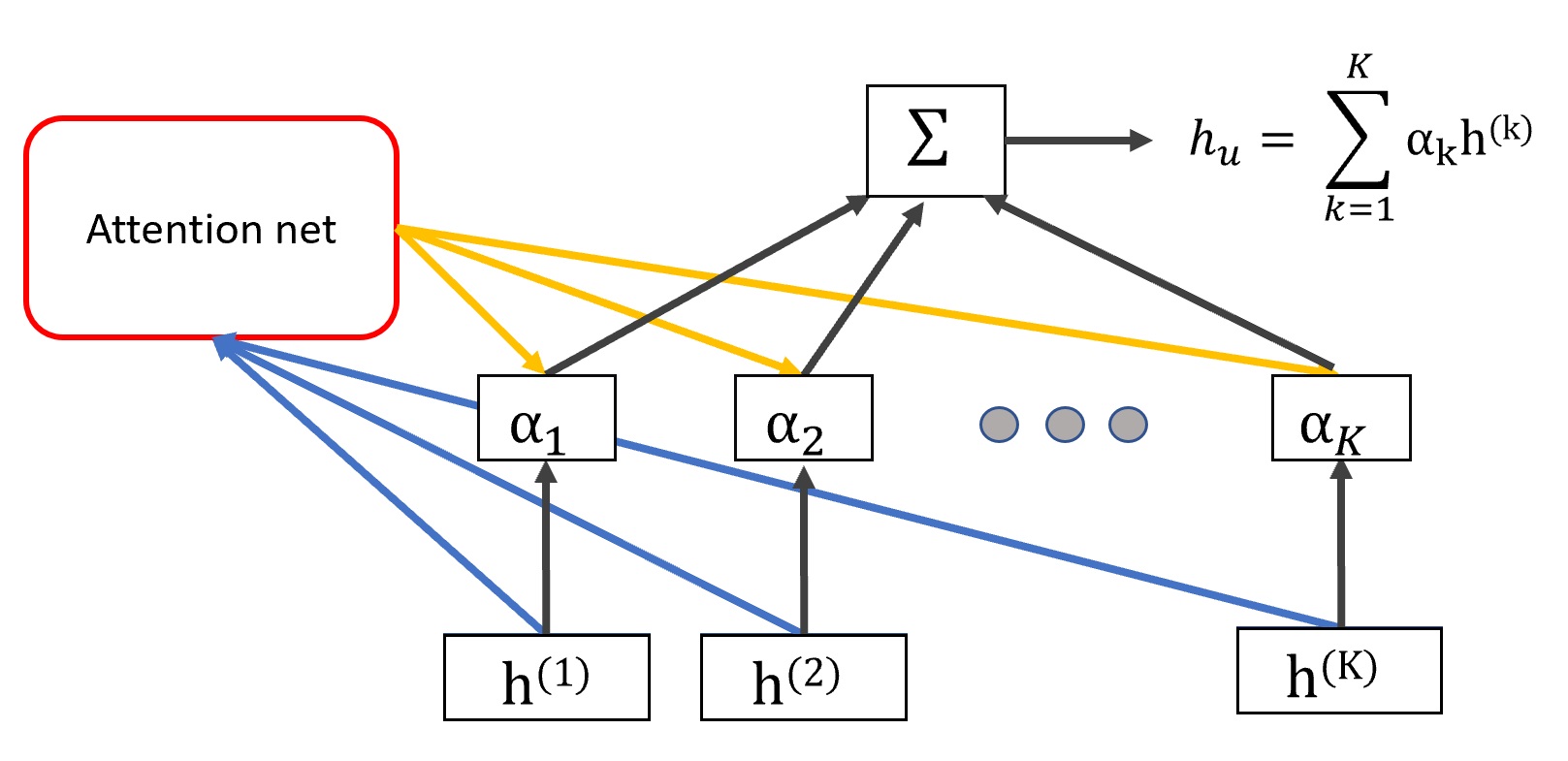}
\caption{Attention-based pooling to combine channel states into user state.}
\vspace{-0.1in}
\label{fig:attention}
\end{figure}

To address this limitation, we propose applying an attention-based mechanism \cite{Bahdanau:2014} as an adaptive pooling strategy. More specifically, the user vector becomes a weighted sum of the channel vectors, where the weights \emph{adapt to the content} of each channel based on intent importance:

\vspace{-0.04in}
\begin{equation}
h_u = \sum_{k=1}^{K} \alpha_k h^{(k)}
\label{eq:attention}
\end{equation}
\vspace{-0.04in}

An additional two-layer feedforward neural network, called the attention network $g(.;\theta)$, is used for learning the $\alpha_k$ weights. This network takes as input the representation of each channel and outputs normalized weights via softmax activation (see Figure \ref{fig:attention}):

\vspace{-0.04in}
\begin{equation}
\alpha_k = \frac{\text{exp}\left(g(h^{(k)}; \theta)\right)} {\sum_{j=1}^{K}\text{exp}\left(g(h^{(j)}; \theta)\right)}
\label{eq:attention-weights}
\end{equation}
\vspace{-0.04in}

Note that this formulation is different from the application of attention in \cite{Zhou:2018, Ni:2018, Zhou:2019:DIEN, Yu:2019:long-short} because our attention weights do not condition on the candidate ad to score at runtime. This allows us to run the user encoder model in a batch mode to continuously update user vectors, which are then consumed for runtime scoring with minimal latency (Section \ref{sec:system} provides more details).

\vspace{-0.02in}
\subsection{Model Training}
\label{sec:learning-formulation}

We propose to utilize data from \emph{search} advertising logs for learning ad recommendation in the queryless \emph{native} advertising scenario. More specifically, we assume that ad $a$ is relevant to user $u$ at a specific time if the $u$ performed a search, clicked on $a$, and then followed up with a conversion on the advertiser website (such as purchasing a specific product or adding to the cart). We use clicks with conversions for defining relevant user-ad pairs is because they imply a stronger shopping intent compared to just clicks. This helps promoting high-quality clicks (clicks that have value to advertisers) while demoting clicks on deceptive ads (``clickbaits'') \cite{Yi:2014}. 

By using this learning formulation as opposed to learning directly from native advertising logs, we attempt to learn the actual user actions as opposed to learning the user \emph{reactions} to the existing native advertising models. This helps in mitigating the data selection bias problem, which is commonly encountered when learning recommendation from biased explicit feedback \cite{schnabel2012}.

Having defined our positive set, we can pose the learning problem as extreme multi-class classification, where the task becomes to classify a specific ad $a$ that user $u$ clicked and converted on among millions of other ads. 

\vspace{-0.06in}
$$P(y=a| u_t) = \frac{\text{exp} \left( S(h_{u(t)}, h_a) \right) }{\sum_{a' \in A} \text{exp} \left( S(h_{u(t)}, h_{a'}) \right) }$$
\vspace{-0.04in}

However, this softmax formulation is impractical because the cost of computing the gradient of the log-likelihood for a single example is proportional to the total number of ads. Furthermore, it assumes that the universe of all possible ads is known a priori and fixed. 

To overcome such difficulties, we encode target ads from the text description and train the model using negative sampling, a technique originally proposed in \cite{Mikolov:2013} for efficiently learning word representations. Accordingly, the task becomes to distinguish the relevant ad from $k$ negative candidates (sampled from the background distribution) using the following formulation:

\vspace{-0.06in}
\begin{equation}
\text{log } \sigma \left( S(h_{u(t)}, h_a) \right) + \sum_{i=1}^k E_{a' \sim P(A)} \left[ \text{log } \sigma \left(- S(h_{u(t)}, h_{a'}) \right) \right]
\label{eq:negative-sampling}
\end{equation}
\vspace{-0.04in}

\noindent where $\sigma(x) = \frac{1}{1+\text{exp}(-x)}$, and $\sigma \left( S(h_{u(t)}, h_a) \right)$ defines the probability of positive label for a specific user-ad pair: $P(y=1|u_t, a)$. This objective encourages relevant user-ad pairs to have a high similarity score (to make the probability close to 1), and vice versa for the negative pairs.

Note that the entire MC-SBN model--namely the parameters of each channel-specific GRU, the attention network, and the ad encoder network--is trained end-to-end from logs to maximize objective in Equation \eqref{eq:negative-sampling}.

\vspace{0.03in}
\textbf{Withholding recent information}: It is important to note that during the process of user data creation from search advertising logs, we explicitly exclude user events that are too close to the time of the target ad impression (e.g., by setting the anchor point to $t-\delta$ if $t$ is the timestamp of the ad impression). Otherwise, the most recent query the user has issued to the search engine would already contain most of the signal for differentiating the target ad from the negatives. Consequently, the model could easily exploit this structure and overfit to the surrogate task by always recommending an ad on the native medium based on the most recent search of the user. Note that a similar setup was described in \cite{Covington:2016}, where authors explicitly avoid using the last search query the user issued on YouTube when learning to recommend new content to watch.

\vspace{-0.02in}
\subsection{System Design}
\label{sec:system}

\begin{figure}
\includegraphics[width=3.2in, height=1.9in]{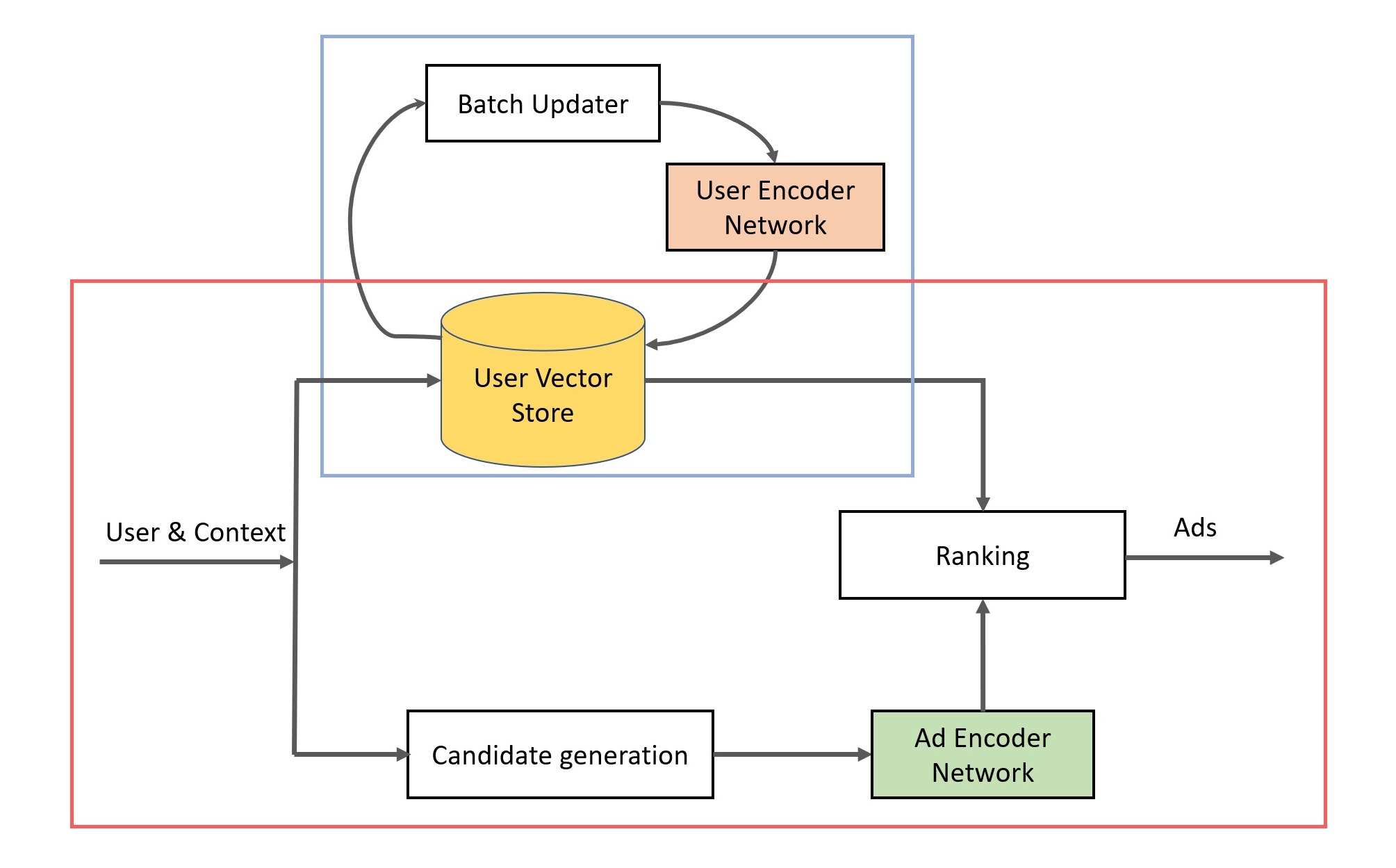}
\vspace{-0.1in}
\caption{System overview. The blue box corresponds to the offline part and the red box corresponds to the online part.}
\label{fig:system}
\end{figure}

In this section, we describe how an already trained model can be efficiently deployed in production. The MC-SBN model is designed to decompose into two sub-models--the user encoder and the ad encoder--which can run independently of each other. The production system has two main components: offline and online (see Figure \ref{fig:system}). The offline system uses only the \emph{user encoder} part of the model and is responsible for maintaining up-to-date user vectors in a key-value store that can be accessed for online serving. An offline batch process--running on a Spark cluster--continuously does the following: (1) read user activities from logs, (2) read existing user vectors from the store, (3) compute the updated user vectors according to Equations \eqref{eq:gru-update}, \eqref{eq:attention} and \eqref{eq:attention-weights}, and (4) write back the updated user vectors.  

At runtime, when a user visits a page which serves native ads, the online system retrieves an initial set of candidate ads and encodes them with the \emph{ad encoder} network\footnote{A caching mechanism is implemented on top to avoid recomputing ad encodings at runtime.}. The system also retrieves from the store the up-to-date vector for the current user. Both user and ad vectors are then leveraged in ranking the ads for a more accurate estimation of relevance, click-through rate, and conversion rate.

\vspace{-0.07in}

\section{Experimental Evaluation}
\label{sec:experiments}

In this section, we evaluate MC-SBN on large-scale datasets obtained from our ad serving system. We define the following two evaluation tasks:
\begin{enumerate}
\item \textbf{Relevance Ranking:} This task aims to identify ads that are relevant to the user by simulating from past serving search advertising logs. 
As discussed in Section \ref{sec:learning-formulation}, we assume that an ad is relevant to a user if the user searches for the ad, clicks on it, and then follows up with a conversion on the advertiser website. The objective is to show that the proposed MC-SBN model outperforms other modeling alternatives in learning to rank relevant ads higher than less relevant ones. 
\vspace{-0.03in}
\item \textbf{Native Ads Prediction Models:} This task studies the impact of incorporating the output of MC-SBN for improving the accuracy of click prediction and conversion prediction in the queryless native advertising setting. 
\end{enumerate}

In the following, we first outline the details of user data (Section \ref{sec:user-channels}). Then we compare the performance of MC-SBN against several other existing approaches for relevance ranking  (Section \ref{sec:relevance-forcasting}). We note some important points about this comparative study and move on to an ablation study to understand gains provided by different components of our model. Lastly, we present results on native advertising data (Section \ref{sec:NA-prediction}).

\vspace{-0.02in}
\subsection{User Data}
\label{sec:user-channels}

We leverage three fundamentally different input channels to represent our users:  

\begin{enumerate}
\item \emph{Page Visits}: This data comes from an event tracking system used by advertisers to report user activity on their websites back to the ad networks. An event--containing the page title, optional keywords, as well as a timestamp on a seconds scale--is sent by the advertiser when a user loads a page or performs a specific action, such as browsing a product. 
\vspace{-0.03in}
\item \emph{Queries}: These are the search queries that users issued to the Bing search engine.
\vspace{-0.03in}
\item \emph{Clicked Ads}: Ads that the user previously clicked on, where a typical ad copy consists of several short sentences describing the ad.

\end{enumerate}

To represent user history, we gather \emph{two weeks} of user data from the three input channels until \emph{one hour} prior to the impression of the target ad. This cut-off at one hour is done to avoid the model memorizing the most recent queries (see Section \ref{sec:learning-formulation}), and also to account for data generation delays which are typical in offline production pipelines. 

\vspace{-0.02in}
\subsection{Relevance Ranking}
\label{sec:relevance-forcasting}

\subsubsection{Dataset}
\label{sec:relevance-forcasting-data}

The dataset, sampled from search advertising logs, is based on the activity of one million unique anonymized users. We use data from $900K$ users for training and $100K$ other users for evaluation. For each user-ad interaction (positive if relevant ad or negative if not), we gather two weeks of user activities prior to the ad as described above. The evaluation task is to infer the true positive out of a set of candidates, which contains the positive and a fixed number of negative samples; as in \cite{Zhou:2018, Yu:2019:long-short, Zhou:2019:DIEN}. 

\subsubsection{Baselines} 
\label{sec:baselines}
We compare MC-SBN with several state-of-the-art models that are used for user recommendation problems \cite{Cheng:2016, Qu:2018:PNN,Zhou:2018,Zhou:2019:DIEN}. One of the unifying themes of the baselines is in the way they represent each user and her activities. These models are ID-based, i.e., each user event (or target ad) is assigned a unique ID and the corresponding embedding is learned as part of the model. The difference comes in the way these models utilize the ID-based embeddings to construct user representations to be used for ranking. We briefly describe the models below:

\begin{itemize}
    \item Wide \& Deep \cite{Cheng:2016} consists of two components: its deep model is a three-layer feedforward neural network that takes as input the concatenation of the embedding of target ad and the average embedding of all user events, whereas its wide model is the cross-product of the ad embedding and the average event embeddings. 	
	\vspace{-0.03in}
    \item PNN \cite{Qu:2018:PNN} uses a product layer to capture interactive patterns between the ad and user events, and feeds it as additional input to the network.
    \vspace{-0.03in}
    \item DIN \cite{Zhou:2018} uses an attention mechanism to activate related user activities based on the target ad.
	\vspace{-0.03in}
    \item DIEN \cite{Zhou:2019:DIEN} is an extension of DIN that takes into account the sequential information of the user events via a two layers of GRU. It also introduces an auxiliary task which tries to predict the next event at each step of the RNN. 
    
\end{itemize} 

\subsubsection{Implementation Details}

For the above-mentioned baselines, we provide a distinct ID to each unique user event (from the three input channels) and each unique target ad that is present in the training dataset. To represent user history, we gather his/her event IDs from all channels and arrange them in order of time (some baseline models utilize this sequential information). During evaluation, if we encounter a new event, i.e., out of vocabulary, we represent it by the average embedding of all events that belong to the same channel. The same applies for target ads not encountered in the training data. 

In contrast to these ID-based models, MC-SBN uses the text associated with user events and ads. Here, we only do simple preprocessing on the raw text by replacing words which have a total frequency of less than $3$ by $<$UNK$>$ token. We do not apply other preprocessing techniques like stop-word removal or stemming. We perform daily segmentation to each input channel and represent the channel as a sequence of sessions in the BoW space as described in Section \ref{sec:channel-representation}. 

All compared models, baselines and MC-SBN, use $128$ vectors to represent the user and ad embeddings. All models are trained using the Adam optimizer with a mini batch $B=512$ examples. We train for a maximum of 100K mini-batch updates with an early-stopping criterion based on the AUC of a validation set (extracted from training data).

\subsubsection{Metrics}

We compare the ranking performance of the models using both global and user-specific ranking metrics. For global ranking, we use the area under the ROC curve (AUC) and the area under the precision-recall curve (PR-AUC). For user-specific ranking, we use the following metrics:

\begin{itemize}
\item MRR: The reciprocal rank for a specific user is the inverse rank of the relevant ad in the candidate set, i.e., $\frac{1}{k}$ when the relevant ad is ranked in the $k$-th position. The mean reciprocal rank (MRR) is the average reciprocal ranks over all users.
	\vspace{-0.03in}
\item $\text{Recall}@k$: The probability that the relevant ad is in the top $k$ ranked ads from the candidate set. We report results using both recalls at 1 and 3 (R@1 and R@3).
\end{itemize}

\begin{table}[!h]
\begin{tabular}{|l|c|c|c|c|c|}
\hline
\textbf{Model} & \textbf{AUC} & \textbf{PR-AUC} &\textbf{MRR} & \textbf{R@1} & \textbf{R@3}\\ \hline \hline
Wide \& Deep & 0.552 & 0.071 & 0.215 & 0.084 & 0.180 \\ \hline
PNN & 0.565 & 0.079 & 0.220 & 0.087 & 0.184 \\ \hline
DIN & 0.569 & 0.073 & 0.210 & 0.076 & 0.177 \\ \hline
DIEN & 0.569 & 0.072 & 0.211 & 0.072 & 0.177\\ \hline
MC-SBN & \textbf{0.827} & \textbf{0.352} & \textbf{0.536} & \textbf{0.386} & \textbf{0.603} \\ \hline
\end{tabular}
\caption{Performance of different models on the relevance ranking task.}
\vspace{-0.05in}
\label{tab:baseline-results}
\vspace{-0.1in}
\end{table}

\subsubsection{Results}
\label{sec:relevance-ranking-results}

The performance of the different baselines and MC-SBN is shown in Table \ref{tab:baseline-results}. We see that the baseline models underperform MC-SBN by a wide margin. The main reason is because these are ID-based models, where items (target ads or user events) are assumed to be from a fixed set. This assumption is not a practical in our domain because many items in the evaluation data might not even be present in the training data (hence treated as out-of-vocabulary). Furthermore, there is a significant contribution of tail items. For instance, target ads having less than $5$ occurrences contribute to approximately $43\%$ of the data. Similarly, user events having less than $5$ occurrences contribute to around $51\%$ of the data. This indicates that we may not have enough data to train good embeddings for these tail targets or events in the baseline models. In comparison, MC-SBN leverages textual features to better generalize on those new or tail items based on their similarity with the other frequent items. Another difference is that the baselines represent a user as a single sequence of events (collected from all input channels), whereas MC-SBN processes the events in sessions and model the input of each channel by its own RNN. As shown in the ablation study, these are also important factors for the better performance of MC-SBN.

\subsubsection{Ablation Study}
\label{sec:ablation-study}

In this section, we conduct experiments to demonstrate the benefits provided by the different components of MC-SBN. We compare the following variants, all of which use the text representation of target ads and user events: 

\begin{itemize}

\item \emph{Word-based models:} These models work directly on the word-level without learning semantic embeddings. A user is simply represented as a text document that contains all terms from her activity history\footnote{Note that word-based models are not practical approaches because they require storing all words from user history for online serving. However, we include them in the comparison to demonstrate the benefit of semantic projections.}. Relevance with target ads is then determined using the cosine similarity in the BoW space with a \emph{tf-idf} weighting scheme. 
	\vspace{-0.03in}
\item \emph{Pooling-based models:} These models learn semantic embeddings from text and also model user channels separately. However, they do not consider the sequential order of sessions within a channel. More specifically, they use \emph{three} different two-layer feedforward networks to embed the sessions of each of the three input channels. From this model class, we use two approaches to represent a \emph{specific channel}: Pool-Last and Pool-Max. The former embed only the most recent session of the channel, while the latter embed each session and then applies \emph{max pooling} over all session representations to get the channel representation. Both of these use the \emph{attention aggregation} proposed in Section \ref{sec:attention} to combine the channels. 
	\vspace{-0.03in}
\item \emph{Sequential models:} All of these variants process the sessions of each channel sequentially using a GRU-based RNN (as described in Section \ref{sec:channel-modeling}). The difference is in the channel aggregation technique. Seq-Max and Seq-Avg use max and average pooling, respectively. Seq-Hid applies a single layer network on top of the channel representations. Lastly, MC-SBN uses our proposed attention aggregator.

\end{itemize}

Note that all pooling-based models and sequential models use the same ad encoder architecture consisting of a bi-directional GRU to model the sequence of words from the ad text. The difference lies in the user encoder network.

The ranking performance of the different baselines is shown in Table \ref{tab:search-results}. The results indicate the following:

\begin{itemize}
\item The word-based model (BoW) significantly underperform DL-based models, showing that simple lexical matching is insufficient for this task.
	\vspace{-0.03in}
\item The pooling-based model that use only most recent sessions (Pool-Last) perform worse than the one utilizing full two weeks of user data (Pool-Max). This demonstrates that models benefit from using more history to improve the user representation. 
	\vspace{-0.03in}
\item Sequential models outperform the atemporal pooling-based model (Pool-Max). The reason is because sequential models are able to understand the order in which activity sessions are observed and learn how to properly incorporate the recency aspect for different semantic concepts, whereas pooling model is oblivious to this and treat all sessions in a similar manner. 
	\vspace{-0.03in}
\item The proposed attention-based channel aggregator (used in MC-SBN) outperforms the other channel aggregation techniques (Seq-Max, Seq-Avg and Seq-Hid), demonstrating the importance of learning adaptive weights to combine channels based on intent importance. 
\end{itemize}

\begin{table}[!h]
\begin{tabular}{|l|c|c|c|c|c|}
\hline
\textbf{Model} & \textbf{AUC} & \textbf{PR-AUC} &\textbf{MRR} & \textbf{R@1} & \textbf{R@3}\\ \hline \hline
BoW & 0.735 & 0.351 & 0.51 & 0.382 & 0.534 \\ \hline
Pool-Last &0.769 & 0.294 & 0.475 & 0.333 & 0.52 \\ \hline
Pool-Max & 0.802 & 0.293 & 0.497 & 0.339 & 0.564 \\ \hline
Seq-Max & 0.805 & 0.305 & 0.501 & 0.346 & 0.566 \\ \hline
Seq-Avg & 0.813 & 0.318 & 0.513 & 0.359 & 0.578 \\ \hline
Seq-Hid & 0.809 & 0.312 & 0.504 & 0.348 & 0.569\\ \hline
MC-SBN & \textbf{0.827} & \textbf{0.352} & \textbf{0.536} & \textbf{0.386} & \textbf{0.603} \\ \hline
\end{tabular}
\caption{Performance of different MC-SBN variants on the relevance ranking task.}
\label{tab:search-results}
\vspace{-0.05in}
\end{table}
\vspace{-0.1in}

\vspace{-0.02in}
\subsection{Native Ads Prediction Models}
\label{sec:NA-prediction}

The previous evaluation shows the advantage of MC-SBN for relevance ranking, a task that aligns with the training objective of the model. In this section, we demonstrate that this model, trained on search advertising logs, provides useful a representation for prediction tasks in the queryless native advertising setting.

We evaluate on two important tasks for native ads: \emph{click prediction} and \emph{conversion prediction}. Click prediction estimates the probability that an ad impression will lead to a click, whereas conversion prediction estimates the probability that an ad click will lead to a conversion for the advertiser. Estimating these probabilities accurately is critical for efficient marketplace management, with the goal of optimizing user experience, profitability of advertising, and the ROI (return on investment) of our advertisers\footnote{Conversion prediction plays a critical role for managing advertisers' ROI and it is incorporated for both the ranking of ads and the pricing of clicks.}.

\subsubsection{Dataset}

The data for click prediction contains native ad impressions together with binary labels to indicate whether the user clicked on the ad. We collect around $100M$ impressions from native advertising logs. The data is split into two equal parts disjoint in time, where the earlier period is used for training the click prediction model and the later one is used for evaluation. This date-based partitioning is consistent with the actual production system deployment and ensures that all model knowledge comes from data before the evaluation period. 

The dataset for conversion prediction consists of native ad clicks with labels reflecting whether the click resulted in a conversion. Similarly, a date-based partitioning is used to create the training and evaluation datasets.

\subsubsection{Baselines}

Models that predict the probability of a click (or a post-click conversion) rely on a rich and diverse set of features that cover different aspects of the ad (e.g., campaign ID, advertiser ID, or ad group), the context (e.g., device type, time of day, or location), the user (e.g., demographics, or user segments), and the publisher. They also include engineered statistics reflecting the historical performance of these different dimensions and their cross product using multiple lookback windows. We refer to those features as the \emph{base} features. 

Once an impression (or a click) is represented as a set of features, the final prediction is obtained using a \emph{gradient boosted decision trees} (GBDT) model \cite{Friedman:2002} that is trained to minimize the cross-entropy loss.

The goal of this evaluation is to study the impact of adding the output of MC-SBN to those base features. More specifically, given a pair of user $u$ and ad $a$, we use the user encoding $h_u$, the ad encoding $h_a$, and the similarity score $S(h_u, h_a)$ as additional features in the GBDT model. Note that the MC-SBN model itself is still trained on search advertising logs (as described in Section \ref{sec:learning-formulation}), but only applied on the native ads datasets to compute these additional semantic features for each user-ad pair.

\subsubsection{Metrics}

We evaluate the global ranking performance of the models using AUC and PR-AUC. We also evaluate advertiser-specific ranking using a per-advertiser AUC (abbreviated as Adv-AUC). This metric computes an AUC for each advertiser separately and then reports a weighted average of the AUCs, where the weight factor reflects the size of data coming from the specific advertiser. Note that reporting Adv-AUC is especially important when evaluating conversion prediction models because it highlights the ability of the model in differentiating good-quality clicks from lower-quality clicks when conditioned on the same advertiser.

In addition to studying the ranking performance of the models, it is important to evaluate their accuracy in estimating the underlying click-through rate (or conversion rate). The reason is because these predicted probabilities are used in various optimizations to determine the optimal allocation and pricing of the ads, and hence should have good probabilistic calibration. For this, we use the relative information gain (RIG) score \cite{Yi:2013}.

\subsubsection{Results}

Table \ref{tab:results-pclick} compares the evaluation metrics for the click prediction task. We observe a major lift across all offline metrics after including MC-SBN features, with +4\% for AUC, +9\% for PR-AUC, +5\% for Adv-AUC, and +19\% for RIG. In particular, the significant improvement in RIG demonstrates a more accurate click-through rate estimation after incorporating the semantic MC-SBN features.

\begin{table}[htbp]
\begin{tabular}{|l|c|c|c|c|c|}
\hline
\textbf{Features} & \textbf{AUC} & \textbf{PR-AUC} & \textbf{Adv-AUC} &\textbf{RIG} \\ \hline \hline
Base & 0.742 & 0.199 & 0.702 & 0.108 \\ \hline
Base + MC-SBN & 0.773 & 0.217 & 0.738 & 0.131 \\ \hline
\end{tabular}
\caption{Click prediction on native ads.}
\vspace{-0.05in}
\label{tab:results-pclick}
\end{table}
\vspace{-0.1in}

Table \ref{tab:results-pconv} shows the results for conversion prediction. Overall, we see that MC-SBN brings a slight improvement in AUC, PR-AUC and RIG. But when considering Adv-AUC, we observe a major improvement with +8\% lift. This demonstrates that the proposed semantic features are useful for a better understanding of user interests, which in turn helps the model to differentiate clicks that are valuable to the advertiser from other lower quality clicks for a better optimization of advertisers' ROI.

\begin{table}[htbp]
\begin{tabular}{|l|c|c|c|c|c|}
\hline
\textbf{Features} & \textbf{AUC} & \textbf{PR-AUC} & \textbf{Adv-AUC} &\textbf{RIG}\\ \hline \hline
Base & 0.812 & 0.248 & 0.558 & 0.193\\ \hline
Base + MC-SBN & 0.827 & 0.257 & 0.603 & 0.204 \\ \hline
\end{tabular}
\caption{Conversion prediction on native ads.}
\vspace{-0.05in}
\label{tab:results-pconv}
\end{table}
\vspace{-0.1in}

\vspace{-0.04in}
\section{Conclusion}
\label{sec:conclusion}

This paper presents MC-SBN, a deep learning approach for understanding user intent with application to online advertising. The proposed user encoder network combines multiple heterogenous data sources to create a user intent representation. Events from each input channel are represented as a sequence of sessions, processed by a specific GRU-based RNN, and channels are then combined using an attention mechanism. The user encoder is deployed offline to continuously update user vectors as events are streamed through the system, which are then leveraged at serving time to efficiently score candidate ads. The experimental evaluation shows that the proposed model outperforms several other modeling approaches for ranking relevant ads higher than non-relevant ones. It also shows that incorporating MC-SBN semantic features significantly improves the accuracy of both click prediction and conversion prediction in native advertising settings.

\bibliographystyle{abbrv}

\bibliography{myBib}

\begin{thebibliography}{10}

\bibitem{Bahdanau:2014}
D.~Bahdanau, K.~Cho, and Y.~Bengio.
\newblock Neural machine translation by jointly learning to align and
  translate.
\newblock {\em CoRR}, abs/1409.0473, 2014.

\bibitem{Bansal:2016}
T.~Bansal, D.~Belanger, and A.~McCallum.
\newblock Ask the gru: Multi-task learning for deep text recommendations.
\newblock In {\em Proceedings of the ACM Conference on Recommender Systems},
  2016.

\bibitem{Beutel:2018}
A.~Beutel, P.~Covington, S.~Jain, C.~Xu, J.~Li, V.~Gatto, and E.~H. Chi.
\newblock Latent cross: Making use of context in recurrent recommender systems.
\newblock In {\em Proceedings of the ACM International Conference on Web Search
  and Data Mining}, 2018.

\bibitem{Blei:2003}
D.~M. Blei, A.~Y. Ng, and M.~I. Jordan.
\newblock Latent dirichlet allocation.
\newblock {\em Journal of Maching Learning Research}, 3:993--1022, 2003.

\bibitem{Cheng:2016}
H.-T. Cheng, L.~Koc, J.~Harmsen, T.~Shaked, T.~Chandra, H.~Aradhye,
  G.~Anderson, G.~Corrado, W.~Chai, M.~Ispir, R.~Anil, Z.~Haque, L.~Hong,
  V.~Jain, X.~Liu, and H.~Shah.
\newblock Wide \& deep learning for recommender systems.
\newblock In {\em Proceedings of the 1st Workshop on Deep Learning for
  Recommender Systems}, 2016.

\bibitem{Cho:2014}
K.~Cho, B.~van Merrienboer, a.~Gülçehre, D.~Bahdanau, F.~Bougares,
  H.~Schwenk, and Y.~Bengio.
\newblock Learning phrase representations using rnn encoder-decoder for
  statistical machine translation.
\newblock In {\em EMNLP}, 2014.

\bibitem{Covington:2016}
P.~Covington, J.~Adams, and E.~Sargin.
\newblock Deep neural networks for youtube recommendations.
\newblock In {\em Proceedings of the 10th ACM Conference on Recommender
  Systems}, 2016.

\bibitem{Donkers:2017}
T.~Donkers, B.~Loepp, and J.~Ziegler.
\newblock Sequential user-based recurrent neural network recommendations.
\newblock In {\em Proceedings of the ACM Conference on Recommender Systems},
  2017.

\bibitem{Friedman:2002}
J.~H. Friedman.
\newblock Stochastic gradient boosting.
\newblock {\em Computational Statistics and Data Analysis}, 38(4):367--378,
  2002.

\bibitem{Hidasi:2015}
B.~Hidasi, A.~Karatzoglou, L.~Baltrunas, and D.~Tikk.
\newblock Session-based recommendations with recurrent neural networks.
\newblock {\em CoRR}, abs/1511.06939, 2015.

\bibitem{Hofmann:1999}
T.~Hofmann.
\newblock Probabilistic latent semantic indexing.
\newblock In {\em Proceedings of SIGIR Conference on Research and Development
  in Information Retrieval}, 1999.

\bibitem{Huang:2013}
P.-S. Huang, X.~He, J.~Gao, L.~Deng, A.~Acero, and L.~Heck.
\newblock Learning deep structured semantic models for web search using
  clickthrough data.
\newblock In {\em Proceedings of the 22nd ACM international conference on
  Conference on information and knowledge management}, 2013.

\bibitem{Le:2014}
Q.~Le and T.~Mikolov.
\newblock Distributed representations of sentences and documents.
\newblock In {\em Proceedings of the International Conference on Machine
  Learning}, 2014.

\bibitem{Liu:2005}
T.~Liu, A.~W. Moore, K.~Yang, and A.~G. Gray.
\newblock An investigation of practical approximate nearest neighbor
  algorithms.
\newblock In {\em Advances in Neural Information Processing Systems}, 2005.

\bibitem{Loyola:2017}
P.~Loyola, C.~Liu, and Y.~Hirate.
\newblock Modeling user session and intent with an attention-based
  encoder-decoder architecture.
\newblock In {\em Proceedings of the ACM Conference on Recommender Systems},
  2017.

\bibitem{Mikolov:2013}
T.~Mikolov, I.~Sutskever, K.~Chen, G.~Corrado, and J.~Dean.
\newblock Distributed representations of words and phrases and their
  compositionality.
\newblock In {\em Neural and Information Processing System (NIPS)}, 2013.

\bibitem{Ni:2018}
Y.~Ni, D.~Ou, S.~Liu, X.~Li, W.~Ou, A.~Zeng, and L.~Si.
\newblock Perceive your users in depth: Learning universal user representations
  from multiple e-commerce tasks.
\newblock In {\em Proceedings of ACM SIGKDD International Conference on
  Knowledge Discovery and Data Mining}, 2018.

\bibitem{Okura:2017}
S.~Okura, Y.~Tagami, S.~Ono, and A.~Tajima.
\newblock Embedding-based news recommendation for millions of users.
\newblock In {\em Proceedings of the SIGKDD International Conference on
  Knowledge Discovery and Data Mining}, 2017.

\bibitem{Peters:2018}
M.~E. Peters, M.~Neumann, M.~Iyyer, M.~Gardner, C.~Clark, K.~Lee, and
  L.~Zettlemoyer.
\newblock Deep contextualized word representations.
\newblock In {\em Proc. of NAACL}, 2018.

\bibitem{Qu:2018:PNN}
Y.~Qu, B.~Fang, W.~Zhang, R.~Tang, M.~Niu, H.~Guo, Y.~Yu, and X.~He.
\newblock Product-based neural networks for user response prediction over
  multi-field categorical data.
\newblock {\em ACM Trans. Inf. Syst.}, 37(1):5:1--5:35, Oct. 2018.

\bibitem{schnabel2012}
T.~Schnabel, A.~Swaminathan, A.~Singh, N.~Chandak, and T.~Joachims.
\newblock Recommendations as treatments: Debiasing learning and evaluation.
\newblock In {\em Proceedings of the International Conference on Machine
  Learning}, 2012.

\bibitem{Shen:2014}
Y.~Shen, X.~He, J.~Gao, L.~Deng, and G.~Mesnil.
\newblock A latent semantic model with convolutional-pooling structure for
  information retrieval.
\newblock In {\em Proceedings of the 23rd ACM International Conference on
  Conference on Information and Knowledge Management}, 2014.

\bibitem{Smirnova:2017}
E.~Smirnova and F.~Vasile.
\newblock Contextual sequence modeling for recommendation with recurrent neural
  networks.
\newblock In {\em Proceedings of the 2Nd Workshop on Deep Learning for
  Recommender Systems}, 2017.

\bibitem{Tang:2017}
L.~Tang and E.~Yi~Liu.
\newblock Joint user-entity representation learning for event recommendation in
  social network.
\newblock In {\em Proceedings of IEEE International Conference on Data
  Engineering}, 2017.

\bibitem{Wang:2002}
C.~Wang, P.~Zhang, R.~Choi, and M.~Eredita.
\newblock Understanding consumer attribute toward advertising.
\newblock {\em Information System}, pages 1143--1148, 2002.

\bibitem{Yi:2013}
J.~Yi, Y.~Chen, J.~Li, S.~Sett, and T.~W. Yan.
\newblock Predictive model performance: offline and online evaluations.
\newblock In {\em Proceedings of the ACM SIGKDD International Conference on
  Knowledge Discovery and Data Mining}, 2013.

\bibitem{Yi:2014}
X.~Yi, L.~Hong, E.~Zhong, N.~N. Liu, and S.~Rajan.
\newblock Beyond clicks: Dwell time for personalization.
\newblock In {\em Proceedings of the ACM Conference on Recommender Systems},
  2014.

\bibitem{Yu:2019:long-short}
Z.~Yu, J.~Lian, A.~Mahmoody, G.~Liu, and X.~Xie.
\newblock Adaptive user modeling with long and short-term preferences for
  personalized recommendation.
\newblock In {\em International Joint Conferences on Artificial Intelligence
  (IJCAI)}, 2019.

\bibitem{Zhai:2016}
S.~Zhai, K.-h. Chang, R.~Zhang, and Z.~M. Zhang.
\newblock Deepintent: Learning attentions for online advertising with recurrent
  neural networks.
\newblock In {\em Proceedings of the ACM SIGKDD International Conference on
  Knowledge Discovery and Data Mining}, 2016.

\bibitem{Zhang:2019}
S.~Zhang, L.~Yao, A.~Sun, and Y.~Tay.
\newblock Deep learning based recommender system: A survey and new
  perspectives.
\newblock {\em ACM Comput. Surv.}, 52(1):5:1--5:38, 2019.

\bibitem{Zhou:2019:DIEN}
G.~Zhou, N.~Mou, Y.~Fan, Q.~Pi, W.~Bian, C.~Zhou, X.~Zhu, and K.~Gai.
\newblock Deep interest evolution network for click-through rate prediction.
\newblock {\em Proceedings of the AAAI Conference on Artificial Intelligence},
  33:5941--5948, 07 2019.

\bibitem{Zhou:2018}
G.~Zhou, C.~Song, X.~Zhu, X.~Ma, Y.~Yan, X.~Dai, H.~Zhu, J.~Jin, H.~Li, and
  K.~Gai.
\newblock Deep interest network for click-through rate prediction.
\newblock In {\em Proceedings of the SIGKDD International Conference on
  Knowledge Discovery and Data Mining}, 2018.

\end{thebibliography}

\end{document}